\documentclass[aps,prl,twocolumn,showpacs,groupedaddress]{revtex4}

\usepackage{amsmath,amssymb}
\usepackage{mathpazo}
\usepackage[mathpazo]{flexisym}
\usepackage{breqn}
\usepackage{graphicx,natbib,times}
\usepackage{tensor}

\newcommand\ii{\rm{i}}
\newcommand\be{\textbf{e}}
\newcommand\bff{\textbf{f}}

\newcommand\bh{\textbf{h}}
\newcommand\bk{\textbf{k}}
\newcommand\bp{\textbf{p}}

\newcommand\bu{\textbf{u}}

\newcommand\bx{\textbf{x}}

\begin{document}
  
\title{Non-Kolmogorov cascade of helicity driven turbulence}

\author {Mouloud Kessar$^{1}$,  Franck Plunian$^{2}$, Rodion Stepanov$^{3,4}$, Guillaume Balarac$^1$\\
$^1$ Univ. Grenoble Alpes, CNRS, LEGI, Grenoble, France;\\ 
$^2$ Univ. Grenoble Alpes, CNRS, ISTerre, Grenoble, France;\\
$^3$ Institute of Continuous Media Mechanics, Korolyov str. 1, 614013 Perm, Russia;\\
$^4$ Perm National Research Polytechnic University, Komsomolskii av. 29, 614990 Perm, Russia}
\email{Franck.Plunian@ujf-grenoble.fr}
\date{\today}
  
\begin{abstract}
We solve the Navier-Stokes equations with two simultaneous forcings. One forcing is applied at a given large-scale and it injects energy. The other forcing is applied at all scales belonging to the inertial range and it injects helicity. In this way we can vary the degree of turbulence helicity from non helical to maximally helical.   
We find that increasing 
the rate of helicity injection does not change the energy flux.
On the other hand the level of total energy is strongly increased and the energy spectrum gets steeper. The energy spectrum spans from a Kolmogorov scaling law $k^{-5/3}$ for a non-helical turbulence, to a non-Kolmogorov scaling law $k^{-7/3}$ for a maximally helical turbulence. In the later case we find that the characteristic time of the turbulence is not the turnover time but a time based on the helicity injection rate.

We also analyse the results
in terms of helical modes decomposition. For a maximally helical turbulence one type of helical mode is found to be much more energetic than the other one, by several orders of magnitude.
The energy cascade of the most energetic type of helical mode results from the sum of two fluxes.
One flux is negative and can be understood in terms of a decimated model. This negative flux is however not sufficient to lead an inverse energy cascade.
Indeed the other flux involving the least energetic type of helical mode is positive and the largest. The least energetic type of helical mode is then essential and cannot be neglected.
\end{abstract}

\maketitle

Considering the case of three-dimensional homogeneous and isotropic turbulence, 
Kolmogorov \cite{Kolmogorov1941} assumed the existence of a range of scales, the so-called inertial range, in which the viscous dissipation can be neglected. 
In absence of dissipation the kinetic energy is a conserved quantity.
In spectral space the flux of kinetic energy is constant, leading to an energy cascade from large to small scales, provided energy is injected at large scale. From a straightforward dimensional analysis the spectral density of kinetic energy $E(k)$ can be expressed in terms of the energy injection rate $\varepsilon$ and wave number $k$,
\begin{equation}
E(k) \propto \varepsilon^{2/3} k^{-5/3}.
\label{scalingE}
\end{equation}

In absence of viscosity
not only energy is a conserved quantity but also helicity \cite{Moreau1961, Moffatt1969}, which is defined as
\begin{equation}
H(t)=\int_V \bu (\bx,t)\cdot\nabla\times\bu(\bx,t) \; dV,
\end{equation}
where $\bu (\bx,t)$ is the velocity field at position $\bx$ and time $t$, and integration is made over the volume $V$.
Similarly to energy, the helicity conservation is equivalent having a constant helicity flux.
Provided energy and helicity are both injected at large scale, helicity is expected to cascade jointly with energy in the inertial range, obeying the following scaling law \cite{Lesieur1971,Brissaud1973,Chen2003b}
\begin{equation}
H(k)\propto \eta \varepsilon^{-1/3} k^{-5/3}
\label{scalingH}
\end{equation}
where $H(k)$ is the spectral density of helicity and $\eta$ the injection rate of helicity.
Contrary to enstrophy in two-dimensional turbulence, helicity is not sign-defined and therefore not reputed for having any influence on the energy spectrum, letting the scaling law (\ref{scalingE}) unchanged \cite{Mininni2006}.

Though the simultaneous scaling laws (\ref{scalingE}) and (\ref{scalingH}) are characteristic of the so-called helical turbulence \cite{Chen2003b}, they cannot be justified from dimensional grounds like Kolmogorov did for the non helical turbulence. Indeed  
the problem now consists in five variables $E(k), H(k), k, \varepsilon$ and $\eta$ and only two dimensions, length and time.
Applying the $\Pi$-theorem \cite{Barenblatt1987} and assuming that $E(k)$ and $H(k)$ obey some scaling laws, we find \cite{Golbraikh2002, Golbraikh2006}
\begin{equation}
E(k) \propto \varepsilon^{7/3-a} \eta^{a-5/3} k^{-a}, \quad  H(k) \propto \varepsilon^{4/3-b} \eta^{b-2/3} k^{-b}
\label{scalingadim}
\end{equation}
where $a$ and $b$ are two free parameters. Therefore we need additional constraints to derive the power laws for $E(k)$ and $H(k)$. 

One way to argue for the simultaneous $k^{-5/3}$ scaling laws (\ref{scalingE}) and (\ref{scalingH})
is to assume that the fluxes of energy and helicity $\Pi_E(k)$ and $\Pi_H(k)$
are constant in the inertial range, such that
$\Pi_E(k)=\varepsilon$ and $\Pi_H(k)=\eta$. In addition we have to set
that the characteristic times $\tau_E$ and $\tau_H$ for the energy and helicity transfers are given by the 
turbulence turnover time  $\tau_E=\tau_H \propto(\varepsilon k^2)^{-1/3}$.
Then estimating both energy and helicity fluxes as \cite{Kraichnan1971,Brissaud1973}
\begin{equation}
\Pi_E(k)=kE(k)/\tau_E(k), \quad \Pi_H(k)= kH(k)/\tau_H(k),
\label{times}
\end{equation}
leads to (\ref{scalingE}) and (\ref{scalingH}).
In the notations of (\ref{scalingadim}) this would correspond to $a=b\equiv 5/3$.

Instead we could think of 
spectral laws independent of $\varepsilon$, leading to \cite{Brissaud1973}
\begin{equation}
E(k)\propto \eta^{2/3}k^{-7/3}, \quad  H(k)\propto \eta^{2/3} k^{-4/3}.
\label{scalings2}
\end{equation}
In the notations of (\ref{scalingadim}) this would correspond to $a=b+1\equiv 7/3$. 
Providing evidence of such scaling laws (\ref{scalings2}), is still a challenging issue and has never been observed so far in direct numerical simulations.  
Recently, a step forward has been made by solving the so-called decimated Navier-Stokes (NS) equations \cite{Biferale2012,Biferale2013}. It consists in splitting each Fourier modes of the velocity field in positive and negative helical modes, and in solving the NS equations keeping only one type of mode.
By construction the resulting turbulence is then exactly maximally helical, i.e. $|H(k)|= k E(k)$ .
In such a decimated model helicity is still a conserved quantity, but now it gets the property to be sign-definite.
It then plays a role similar to enstrophy in 2D turbulence, leading to a $k^{-5/3}$ inverse cascade of energy at scales larger than the forcing scale \cite{Biferale2012}. In the inertial range the helicity cascade is direct, with an energy scaling law $E(k)\propto k^{-7/3}$ \cite{Biferale2013}. 
In a recent experimental study \cite{Herbert2012} two scaling laws have also been found, the authors arguing for the existence of two such opposite cascades, but with dominant non-local transfers leading to $E(k)\propto k^{-1}$ and $E(k)\propto k^{-2}$.

Here we present another strategy that does not assume any simplification of the NS equations and which is based on the fact that the scaling laws (\ref{scalings2}) do not depend on $\varepsilon$.
Such a $\varepsilon$ independence is expected as soon as $\tau_H\le\tau_E$, then speaking of a helicity driven turbulence. From (\ref{times}) and applying the exact constraint 
$|H(k)|\le k E(k)$, a sufficient condition for having $\tau_H\le\tau_E$ is given by
\begin{equation}
\Pi_H(k) \ge k \Pi_E(k).
\label{fluxcondition}
\end{equation}
One way to satisfy such a flux condition (\ref{fluxcondition})
is to inject energy at large scale such that $\Pi_E(k)=\varepsilon$ and helicity at all scales such that $\Pi_H(k) \ge k \varepsilon$. 
This is the strategy that is followed in the present letter using direct numerical simulation of the NS equations. 
A similar strategy has been used in \cite{Stepanov2014,Stepanov2015} using a helical shell model.

Using a pseudo-spectral code we solve the NS equations
\begin{equation}
\partial_t\bu = -(\bu\cdot\nabla)\bu -\nabla p + \nu\nabla^2\bu + \bff
\label{NS}
\end{equation}
where $\nu$, $p$ and $\bff$ are respectively the viscosity, the pressure and the flow forcing.
The forcing is divided in two parts $\bff=\bff^E+\bff^H$, where $\bff^E$
is the energy forcing applied at some given large scale $k_F^{-1}$, and $\bff^H$ is the helicity forcing applied at all scales within the inertial range. 

Both parts of the forcing $\bff^E$ and $\bff^H$ are delta-correlated in time and divergence free.
Following \cite{Alvelius1999} they are defined such that the power input comes from the force-force correlation only and not from the velocity-force correlation.
In spectral space this corresponds to
\begin{eqnarray}
\bu_k^*\cdot\bff^E_k + c.c.&=& 0\\
 \bu_k^*\cdot\bff^H_k + c.c.&=& 0 \label{E=0_alv}
\end{eqnarray}
where $\bff^E_k$ and $\bff^H_k$ are the Fourier coefficients of $\bff^E$ and $\bff^H$.

For $\bff^E$ we use the exact same forcing as in \cite{Alvelius1999} with a force-force correlation given by
\begin{eqnarray}
|\bff^E_k|^2&=&F(\bk) / 2\pi k^2, \label{P_alv}
\end{eqnarray}
where $F(\bk)$ obeys to a Gaussian distribution around $k=k_F$. 
As in \cite{Alvelius1999} $F(\bk)$ is defined as inversely  proportional to the time step of the computation,
in order to guarantee an injection rate of energy which is independent from the value of the time step.
The level of helicity injected by $\bff^E$ is not controlled a priori, but the results show that it is statistically  insignificant.

In order to inject helicity the forcing $\bff^H$ has to satisfy, in spectral space,
\begin{eqnarray}
(\nabla \times \bu)_k^* \cdot \bff^H_k + \bu_k^* \cdot (\nabla \times \bff^H)_k + c.c.&=&\eta(\bk), \label{H=eta0}
\end{eqnarray}
where $\eta(\bk)$ is a helicity injection rate per unit volume. 
We take 
\begin{eqnarray}
\eta(\bk)=&0&\text{for } |\bk|< k_F \\
\eta(\bk)=&\eta_0 (|\bk|/k_F)^{-\alpha}&\text{for } |\bk|\ge k_F
\end{eqnarray}
with $\alpha=2.2$
in order to have a spectral density of helicity injection rate $|\bk|^2\eta(\bk)$ almost flat.
 Of course such a forcing extending on the whole inertial range might change the intermittency properties of the turbulence \cite{Biferale2004b}. However we find that the level of dissipation with and without $\bff^H$ is statistically unchanged. 
Finally two issues have to be clarified, both related to the fact that the energy power coming from the force-force correlation of $\bff^H$ is not controlled a priori and that we need to keep it at a level sufficiently lower than the one injected by $\bff^E$.
These rather technical issues are detailed in the Appendix.

Applying a classic criterion \cite{Pope2000} in order to ensure the resolution of a sufficiently large range of dissipation scales, 
taking a grid of $256^3$ points
and setting $\nu=2\;10^{-3}$ and $R_{\lambda}=100$, where $R_{\lambda}$ is 
the Reynolds number based on the Taylor microscale, leads to an energy injection rate $\varepsilon \approx 0.2$
and a forcing wave number $k_F \approx 2.2$.
Finally all subsequent results correspond to statistically steady states.

\begin{figure}
\includegraphics[angle=-90,width=0.4\textwidth]{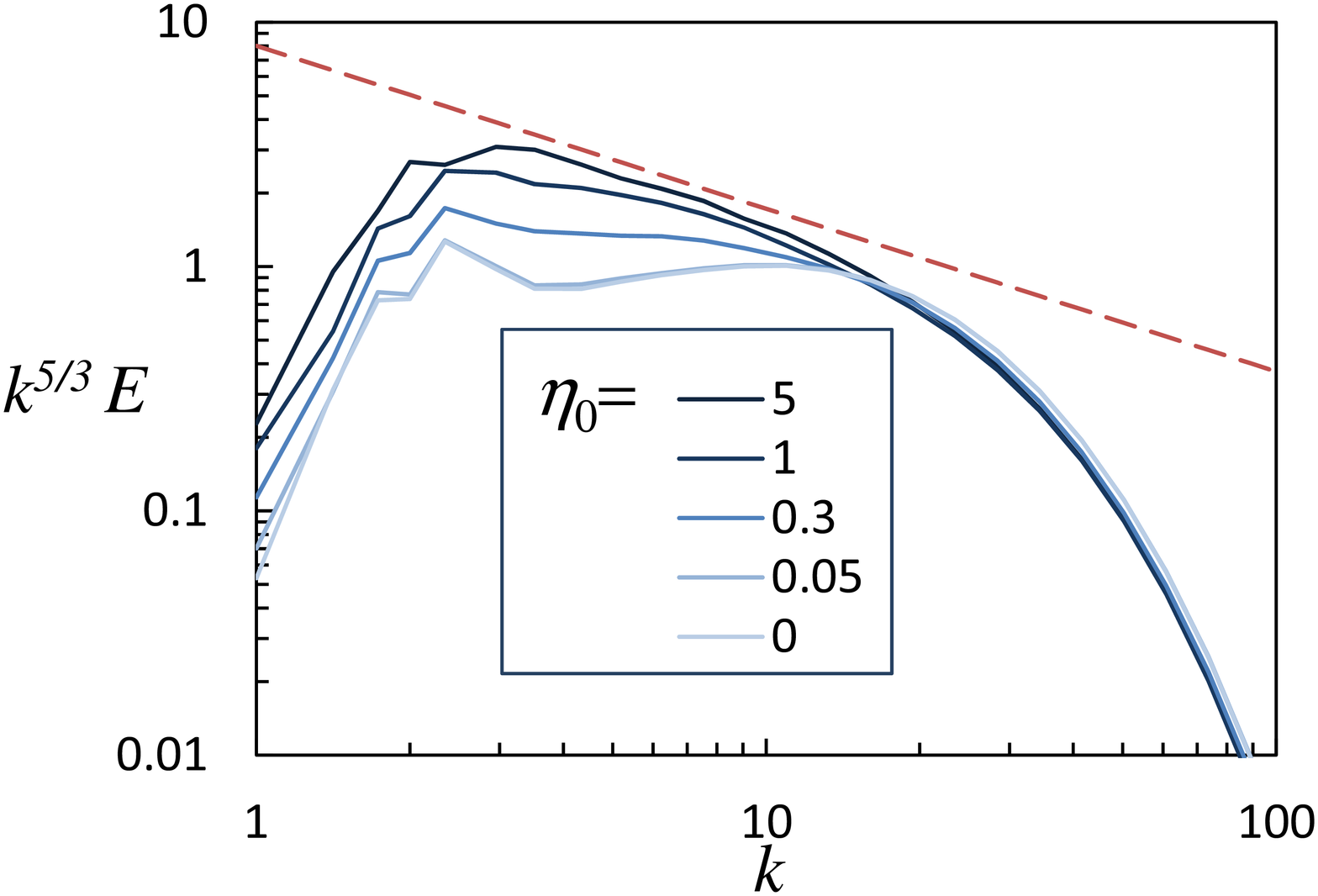}
\includegraphics[angle=-90,width=0.4\textwidth]{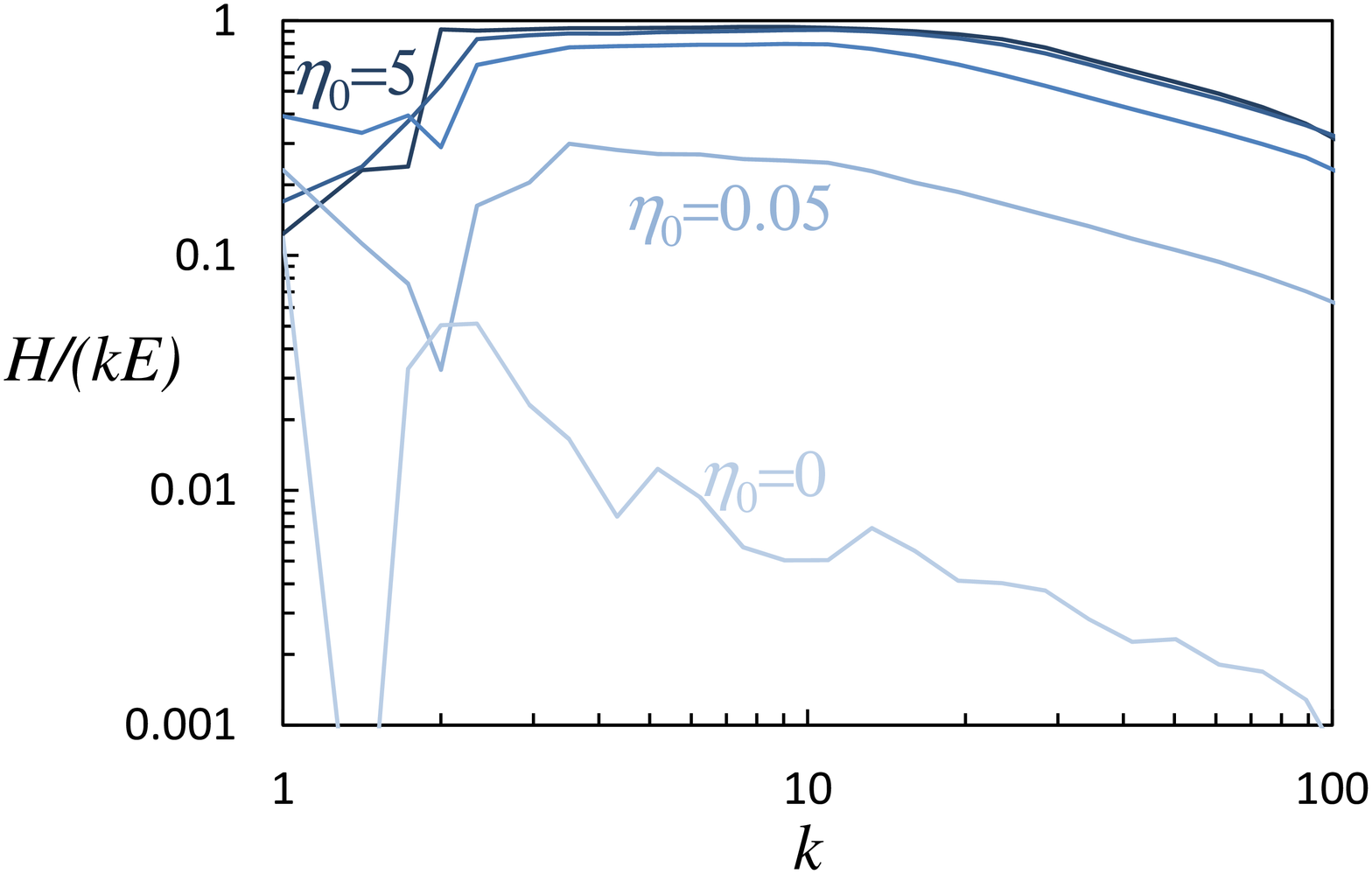}
\caption{(Color online) Spectral density of energy (top) and relative helicity (bottom), for five values of the helicity injection rate: $\eta_0=0, 0.05, 0.3, 1$ and $5$. In the top figure the energy is normalized by $k^{-5/3}$ and the red dashed curve corresponds to $k^{-7/3}$.}
\label{figspectra}
\end{figure}

In Figure \ref{figspectra} top and bottom, the spectral density of energy $E(k)$ and
relative helicity $H(k)/(kE(k))$ are represented for five values of helicity injection $\eta_0$, ranging
from non-helical turbulence ($\eta_0=0$) to maximally helical turbulence ($\eta_0=5$).
Clearly, increasing $\eta_0$ steepens the energy spectral density at large scales, with a scaling law varying from $k^{-5/3}$ for non-helical, to $k^{-7/3}$ for maximally helical turbulence (top of Figure \ref{figspectra}).
For $\eta_0 \neq 0$ a well defined spectrum of relative helicity is obtained with a rather flat part.  For $\eta_0=1$ and $\eta_0=5$ the relative helicity is about unity over
an extended range of scales, showing that the turbulence is close to a maximally helical state (bottom of Figure \ref{figspectra}).

\begin{figure}
\centering
\includegraphics[angle=-90,width=0.4\textwidth]{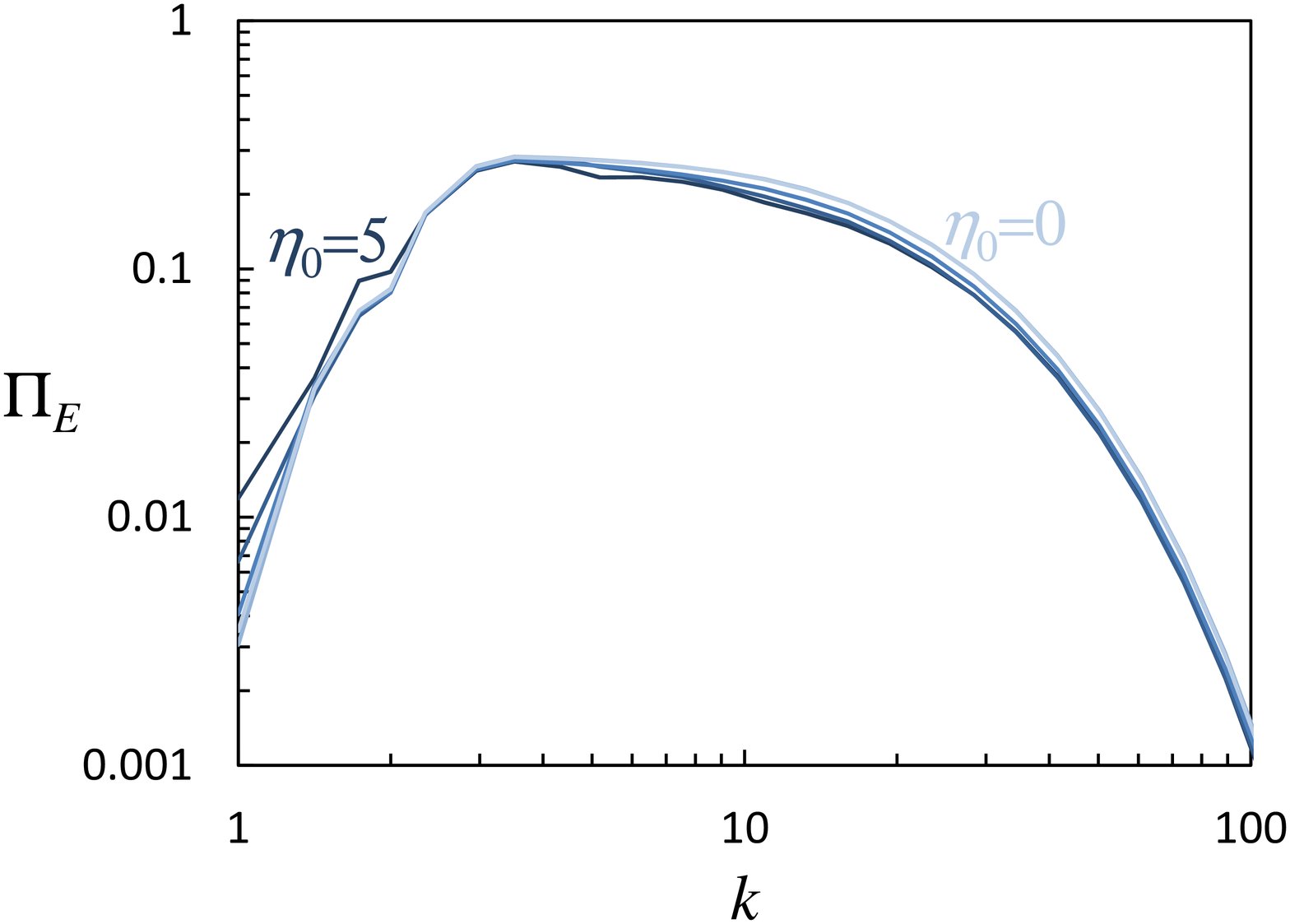}
\includegraphics[angle=-90,width=0.4\textwidth]{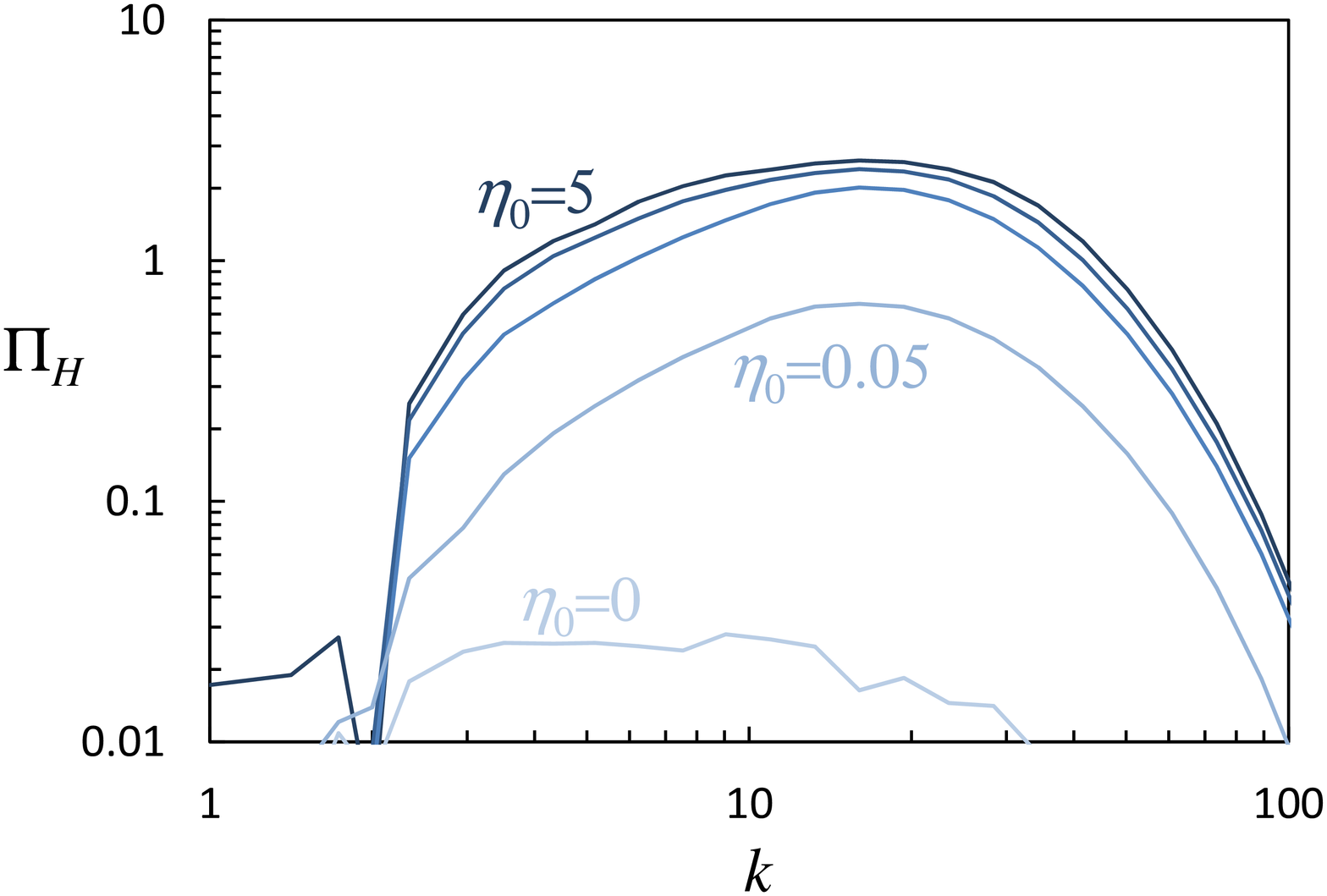}
\caption{(Color online) Flux of energy $\Pi_E$ (top) and helicity $\Pi_H$ (bottom) for the same five values of $\eta_0$ as in Figure \ref{figspectra} and same color code.}
\label{figfluxes}
\end{figure}

The fluxes of energy and helicity, $\Pi_E(k)$ and $\Pi_H(k)$, are plotted in top and bottom of Figure \ref{figfluxes}, for again the same five values of $\eta_0$. In the top figure we see that $\Pi_E(k)$ is almost independent of $\eta_0$,
showing that the spurious energy injection produced by the helical forcing $\bff^H$ is small compared to the energy injected by $\bff^E$. 
On the other hand in the bottom figure we see that $\Pi_H(k)$ is getting higher when increasing the value of $\eta_0$.
Therefore we conclude a posteriori that the injections of energy and helicity are well prescribed by $\bff^E$ and $\bff^H$ respectively. 

\begin{figure}
\centering
\includegraphics[angle=-90,width=0.4\textwidth]{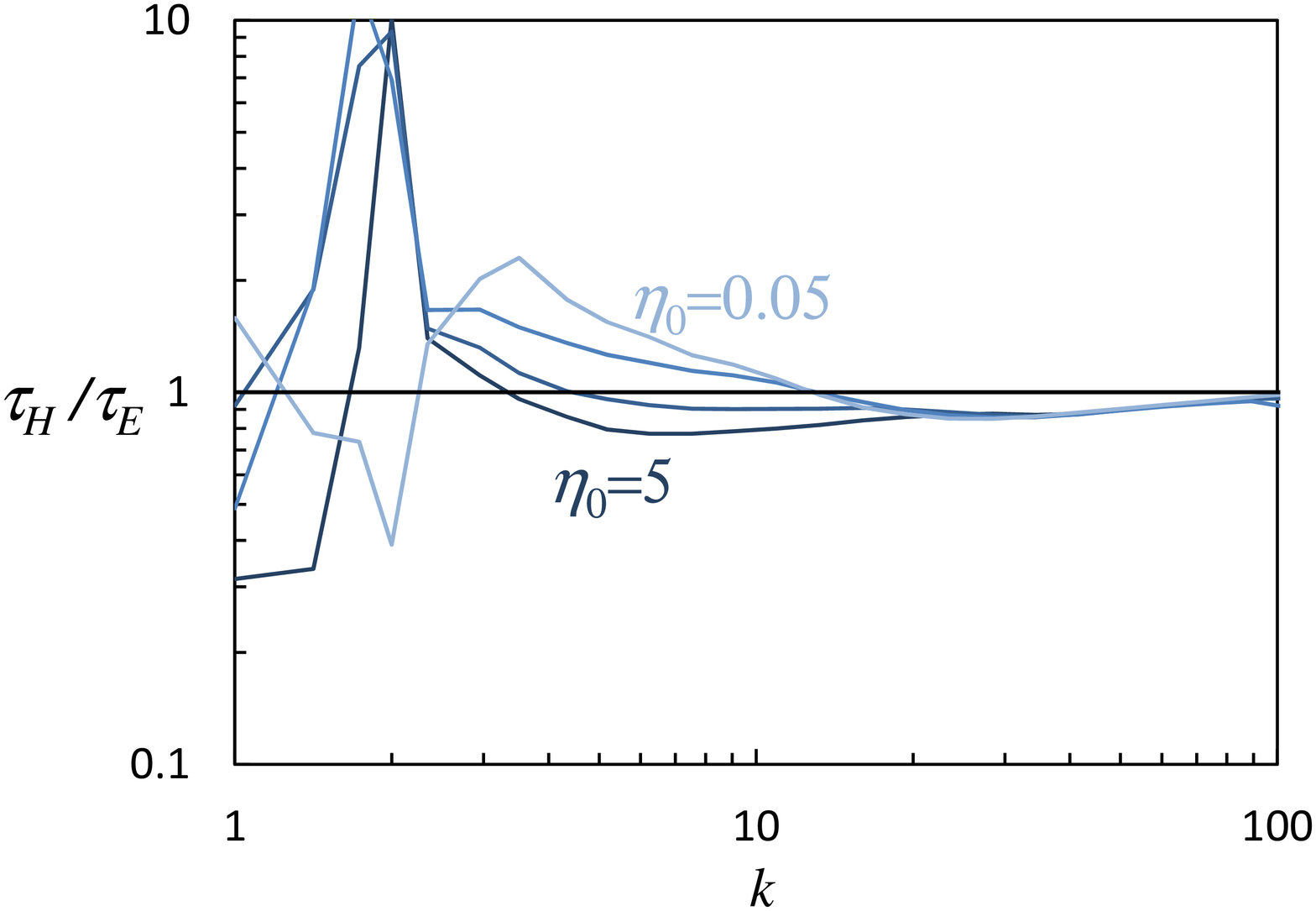}
\caption{(Color online) Ratio $\tau_H(k) /\tau_E(k)$ versus $k$ for $\eta_0=0.05, 0.3, 1$ and $5$ and same color code as in Figure \ref{figspectra}.}
\label{figtimeratio}
\end{figure}

Relying on (\ref{times}) and knowing the flux and spectral density of energy and helicity,  we can calculate the two characteristic times, $\tau_E(k)$ and $\tau_H(k)$, in order to determine which one is the smallest and therefore which one controls the turbulence.
In Figure \ref{figtimeratio} the ratio $\tau_H(k) /\tau_E(k)$ is plotted for $\eta_0= 0.05, 0.3, 1$ and $5$.
For sufficiently large values of $\eta_0$, typically $\eta_0=1$ and $\eta_0=5$, we see that for $k \in [3,12]$ $\tau_H(k)/\tau_E(k)<1$, suggesting a turbulence governed by the helicity injection rate. On the other hand for low values of $\eta_0$, typically $\eta_0=0.05$ and $\eta_0=0.3$, and in the same range of scales we find $\tau_H(k)/\tau_E(k)>1$, suggesting a turbulence governed by the energy injection.

Up to now it has been demonstrated that injecting a sufficiently high rate of helicity over the whole inertial range of a turbulent flow leads to a $k^{-7/3}$ scaling law for the energy spectral density and that the characteristic time of such maximally helical turbulence is the one based on the helicity injection rate. Apart from studies assuming drastic simplification of the NS equations \cite{Biferale2012,Biferale2013,Stepanov2014,Stepanov2015}, this is the first direct numerical simulation in which helicity is shown to have some effect on three-dimensional homogeneous and isotropic turbulence. 

Now the question arises how our results fit in with the scenario described by the decimated model \cite{Biferale2012,Biferale2013}. As the injection of positive helicity is made at all scales and at each time step, we expect a strong dominance of the positive helical modes compared to the negative helical modes. Then according to \cite{Biferale2012,Biferale2013} we could expect an inverse cascade of energy. However the energy fluxes plotted in Figure \ref{figfluxes} are always positive, demonstrating a direct cascade of energy. To clarify this paradox we now analyse our results in terms of helical modes decomposition.

\begin{figure}
\centering
\includegraphics[angle=-90,width=0.4\textwidth]{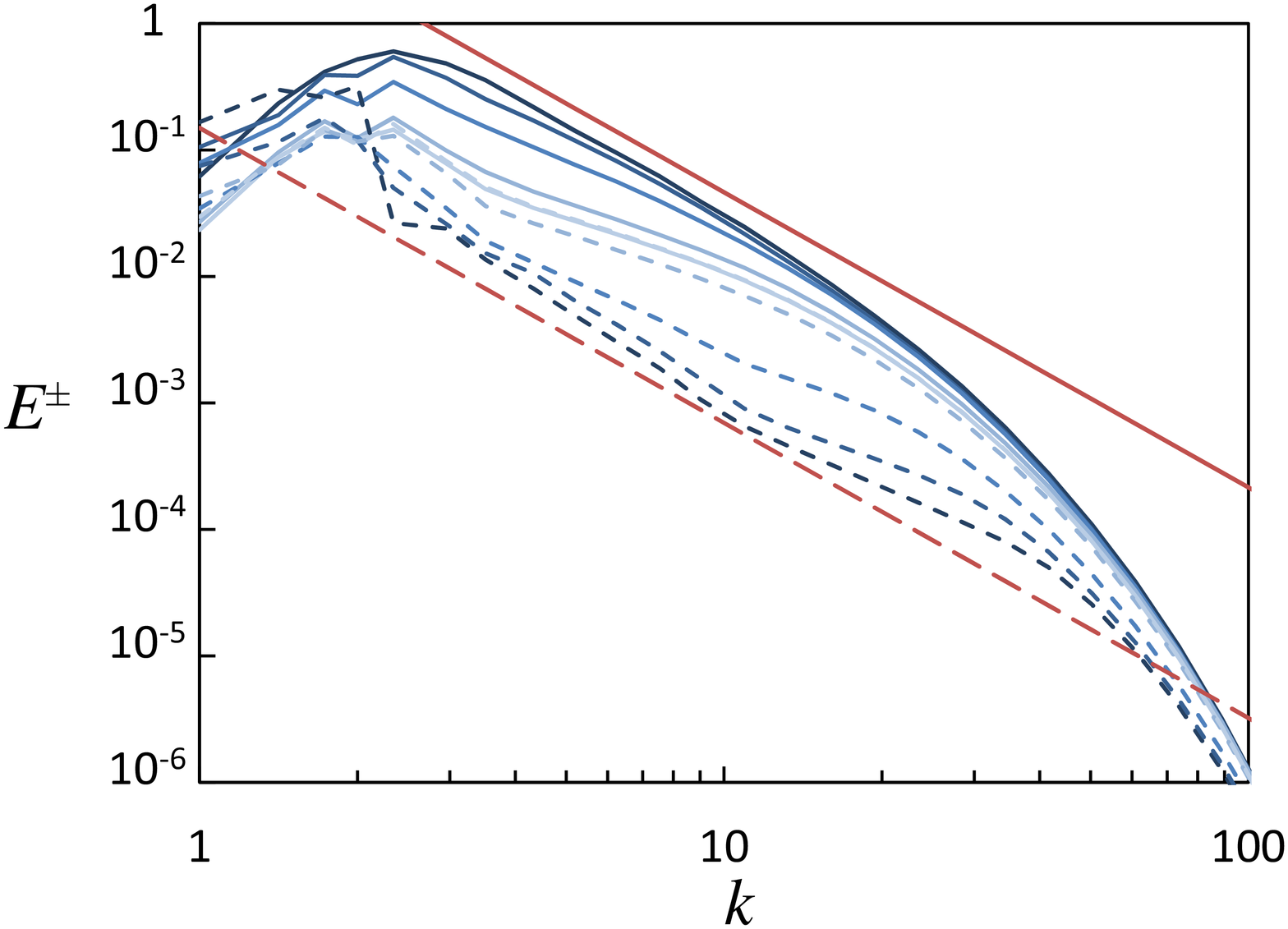}
\caption{(Color online) Energy spectra $E^+(k)$ (solid line) and $E^-(k)$ (dashed line) for the same five values of $\eta_0$ as in Figure \ref{figspectra} and same color code. Increasing $\eta_0$ the curves spread from the center towards the two (red) dashed and solid straight lines corresponding to $k^{-7/3}$.}
\label{energyhelical}
\end{figure}

In Fourier space the velocity field is split into two helical modes per wave vector  
\begin{eqnarray}
\bu(\bk)&=&\bu^+(\bk)+\bu^-(\bk)\\
        &=&u^+(\bk)\bh^+(\bk) + u^-(\bk)\bh^-(\bk)
\label{decomp-mode-hel}
\end{eqnarray}
where $u^{\pm}$ are complex scalars and
$\bh^{\pm}$ are the eigenvectors of the curl operator satisfying $\ii \bk\times\bh^{\pm}=\pm|\bk|\bh^{\pm}$ \cite{Waleffe1992,Lessinnes2011}. 
The energy spectral density of each helical mode, defined by $E^{\pm}(k)=|u^{\pm}(k)|^2$, is plotted in Figure \ref{energyhelical}, for again the same five values of $\eta_0$ as in Figure \ref{figspectra}. 
For each value of $\eta_0$ we observe that 
both spectra $E^{\pm}(k)$ obey to the same scaling laws, again varying from $k^{-5/3}$ for $\eta_0=0$ to $k^{-7/3}$ for $\eta_0=5$. This is consistent with the results of Figure \ref{figspectra} and the relation $E(k)=E^+(k) + E^-(k)$. For $\eta_0=0$ both spectra $E^{\pm}(k)$ are identical as expected for a non helical turbulence. Increasing $\eta_0$ both spectra separate apart from each other, $E^+(k)$ prevailing over $E^-(k)$ by two orders of magnitude for $\eta_0=5$.

\begin{figure}
\includegraphics[angle=-0,width=0.32\textwidth]{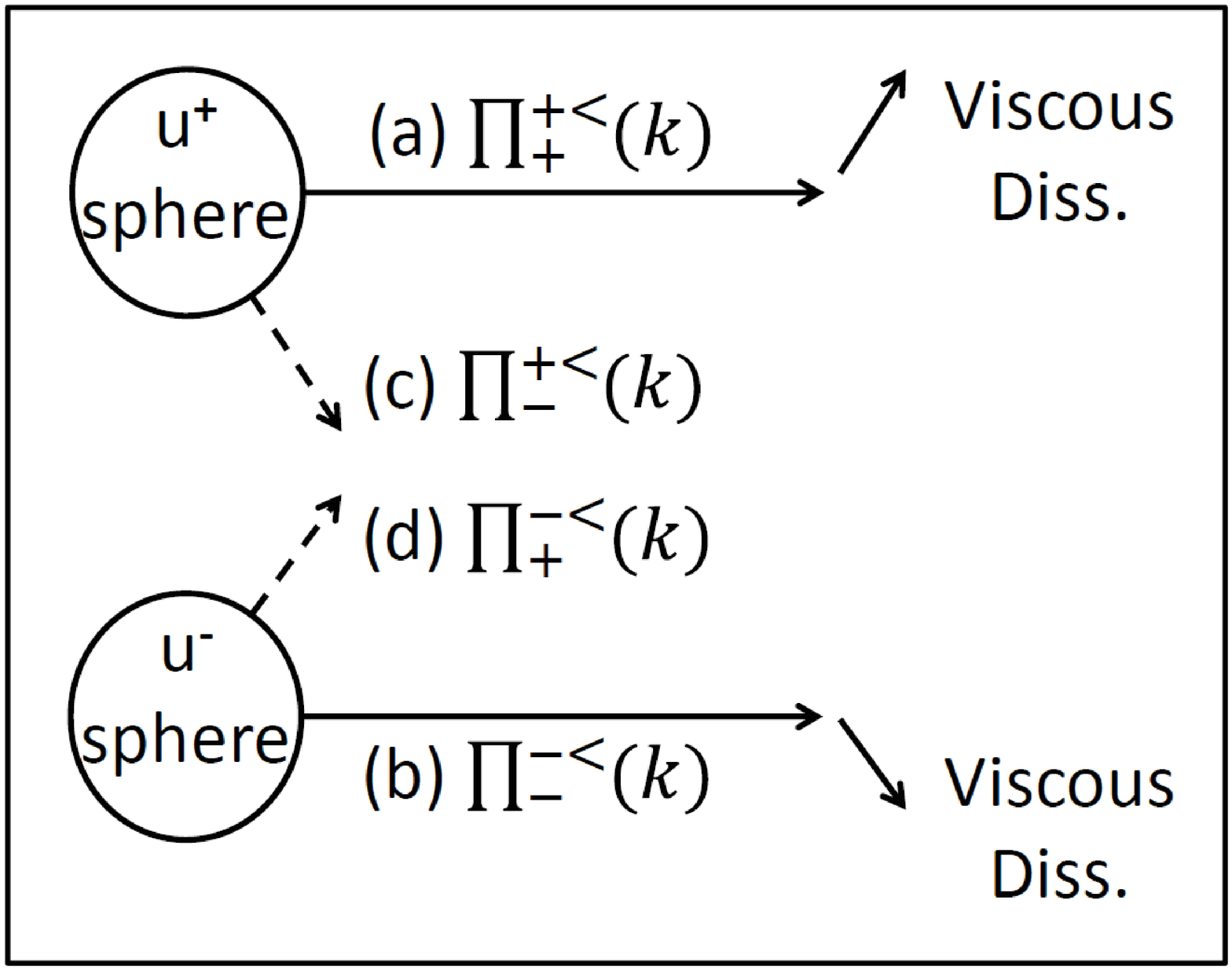}
\includegraphics[angle=-90,width=0.4\textwidth]{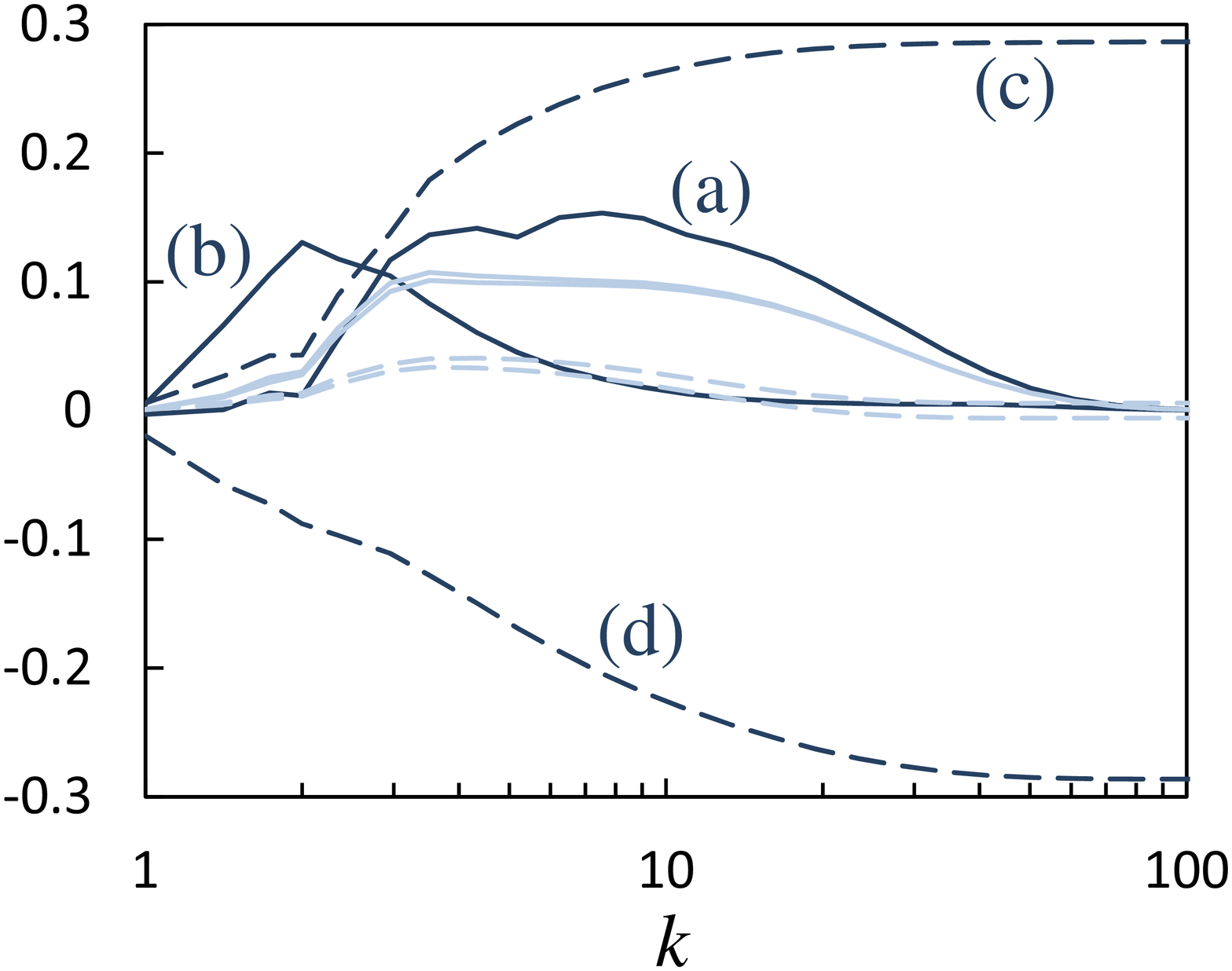}
\caption{(Color online) Top: Various energy fluxes in helical turbulence. $\Pi^{a<}_{b}(k)$ denotes the energy flux from the inside region of a $\bu^{a}$-sphere of radius $k$ to all wave numbers of $\bu^{b}$, where $a,b\equiv\pm$.
Bottom: The dark (light) curves correspond to $\eta_0=5$ ($\eta_0=0$). 
The solid curves correspond to 
$\Pi^{+<}_{+}(k)$ and $\Pi^{-<}_{-}(k)$, the dashed curves to $\Pi^{+<}_{-}(k)$ and $\Pi^{-<}_{+}(k)$.
For $\eta_0=5$: (a) corresponds to $\Pi^{+<}_{+}(k)$, (b) to $\Pi^{-<}_{-}(k)$, (c) to $\Pi^{+<}_{-}(k)$ and (d) to $\Pi^{-<}_{+}(k)$.
}
\label{figfluxeta5}
\end{figure}

Following \cite{Verma2004,Plunian2013} we now analyse the fluxes between the helical modes. We denote by $\Pi^{a<}_{b}(k)$ the energy flux from the inside region of a $\bu^{a}$-sphere of radius $k$ to all wave numbers of $\bu^{b}$, where $a,b\equiv\pm$. It is defined as 
\begin{eqnarray}
\Pi^{a<}_{b}(k)=\int_{|\bk'|\le k}\bu^{a}(\bk')\cdot FT\{(\bu\cdot\nabla)\bu^b\}(\bk')d\bk'
\label{flux}
\end{eqnarray}
where
$FT\{(\bu\cdot\nabla)\bu^b\}$ denotes the Fourier transform of the non linear term $(\bu\cdot\nabla)\bu^b$. 
The four fluxes $\Pi^{\pm<}_{\pm}(k)$ are represented schematically in Figure \ref{figfluxeta5} (top). They are
 plotted in Figure \ref{figfluxeta5} (bottom) for $\eta_0=0$ (light curves) and $\eta_0=5$ (dark curves). The sum of these four fluxes corresponds to the
energy flux plotted in Figure \ref{figfluxes} (top).
For $\eta_0=5$ the fluxes (c) $\Pi^{+<}_{-}(k)$ and (d) $\Pi^{-<}_{+}(k)$ are much larger than for $\eta_0=0$. In addition they are of opposite sign corresponding to a net flux of energy from the positive to the negative helical modes, balancing each other at small scales (large $k$).

Let us now focus on the flux (a) $\Pi^{+<}_{+}(k)$. By definition the energy flux from the inside region of a $\bu^{+}$-sphere of radius $k$ to itself is zero. This implies that $\Pi^{+<}_{+}(k)=\Pi^{+<}_{+>}(k)$, which is the energy flux from the inside region of the $\bu^+$-sphere of radius $k$ to the outside of that same $\bu^+$-sphere. Now the fact that in Figure \ref{figfluxeta5} (bottom) $\Pi^{+<}_{+>}(k)$ is always positive means that there is a direct cascade of energy. This is in contrast with the inverse cascade found with the decimated model of \cite{Biferale2012,Biferale2013}. 

\begin{figure}
\includegraphics[angle=-90,width=0.4\textwidth]{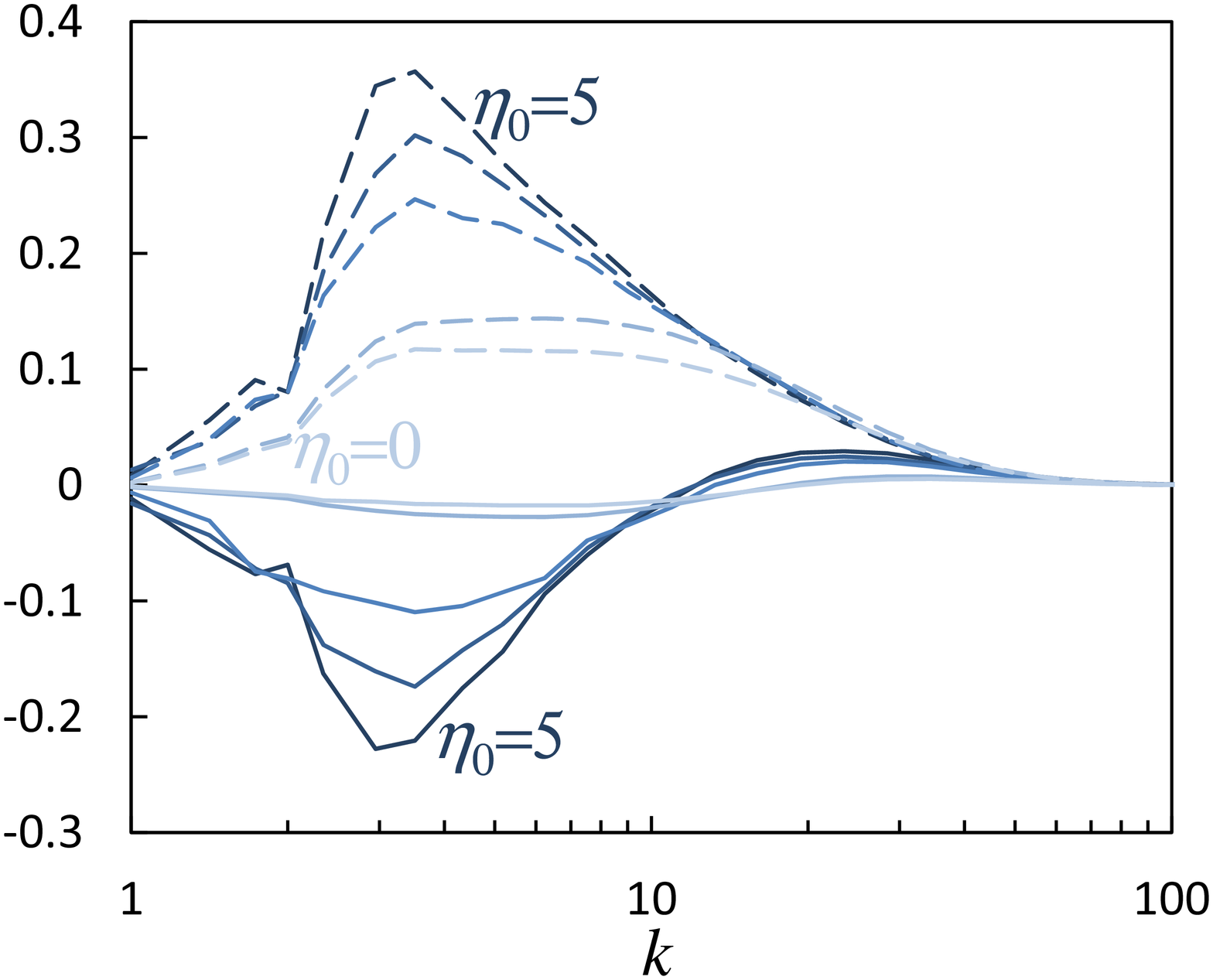}
\caption{(Color online) Energy fluxes ${}^{+}\Pi^{+<}_{+>}(k)$ (solid) and ${}^{-}\Pi^{+<}_{+>}(k)$ (dashed) for the same five values of $\eta_0$ as in Figure \ref{figspectra} and same color code. Increasing $\eta_0$ from 0 to 5 the solid curves at $k=3$ decrease and the dashed curves increase.}
\label{figflux+++eta5}
\end{figure}

Finally we push one step further by splitting the flux 
$\Pi^{+<}_{+>}(k)$ into two parts
$\Pi^{+<}_{+>}(k)={}^+\Pi^{+<}_{+>}(k)+{}^-\Pi^{+<}_{+>}(k)$ with
\begin{eqnarray}
{}^{\pm}\Pi^{+<}_{+>}(k)
=\int_{|\bk'|\le k}\bu^+(\bk')\cdot FT\{(\bu^{\pm}\cdot\nabla)\bu^+\}(\bk')d\bk'.
\label{flux2}
\end{eqnarray}
In (\ref{flux2}) ${}^{\pm}\Pi^{+<}_{+>}(k)$ denotes the energy flux from the inside region of a $\bu^+$-sphere of radius $k$ to the outside of the $\bu^+$-sphere, with $\bu^{\pm}$ acting as a mediator on the non linear interactions. Both fluxes 
${}^{\pm}\Pi^{+<}_{+>}(k)$ are plotted in Figure \ref{figflux+++eta5} for again the same five values of $\eta_0$ as in Figure \ref{figspectra}. The flux ${}^{+}\Pi^{+<}_{+>}(k)$ is always negative in agreement with the arguments given in \cite{Biferale2012,Biferale2013} for the decimated model. However the  flux ${}^{-}\Pi^{+<}_{+>}(k)$ is positive and always the largest in absolute value. This shows that even if the turbulence is strongly positively helical, the presence of negative helical modes is nevertheless essential to give the right sign of the energy fluxes. Though the decimated model is mathematically appealing because it reproduces an exact maximally helical flow, it is eventually singular as in practice the existence of both types of helical modes cannot be avoided. This result also supports the choice made in helical shell models \cite{Plunian2013} in which two helical modes can interact only if they have opposite helicities. 

\begin{acknowledgements}
ISTerre and LEGI are part of Labex OSUG@2020 (ANR10LABX56) and Labex Tec21 (ANR11LABX30). 
We acknowledge support from region Rh\^one-Alpes through the CIBLE program, IDRIS and CIMENT for HPC resources.
We acknowledge provision for computational resources of the URAN
and  TRITON  clusters of Russian Academy of Science, Ural Branch.
\end{acknowledgements}

\section{APPENDIX: the helicity forcing}
\label{app}
The helicity forcing is defined in its spectral form as 
\begin{equation}
\bff^H_k=\bk\times\bp(\bk,t)
\end{equation}
with
\begin{equation}
\bp(\bk,t)=a(\bk,t)\be_1(\bk,t)+ {\ii} b(\bk,t)\be_2(\bk,t),
\end{equation}
$\be_1(\bk,t)$ and $\be_2(\bk,t)$ being two unit vectors with directions changing randomly at each time step.
The resolution of equations (\ref{E=0_alv}) and (\ref{H=eta0}) leads to 
\begin{eqnarray}
a(\bk,t) &= -&\frac{\eta(\bk)}{4D|\bk|^2}\left(\Im{(\bu_k)},\bk,\be_2\right) \label{a}\\
b(\bk,t) &=& \frac{\eta(\bk)}{4D|\bk|^2}\left(\Re{(\bu_k)},\bk,\be_1\right), \label{b}
\end{eqnarray}
with
\begin{eqnarray}
D(\bk,t)&=&\left(\Re{(\bu_k)},\bk,\be_1\right) \left(\Re{(\bu_k)}\cdot\be_2\right)\nonumber\\
&+& \left(\Im{(\bu_k)},\bk,\be_2\right) \left(\Im{(\bu_k)}\cdot\be_1\right).
\label{D}
\end{eqnarray}
As stated above, the energy forcing $\bff^E$ is inversely proportional to the time step \cite{Alvelius1999}.
Conversely $\bff^H$ does not depend on the time step, implying that the level of energy rate which comes from the force-force correlation of $\bff^H$ is proportional to the time step.
Therefore provided the time step is sufficiently small, the energy rate coming from the force-force correlation of $\bff^H$ can be maintained at a sufficiently low level compared to the energy rate coming from the force-force correlation of $\bff^E$. 
In other words to maintain a spurious power injected by $\bff^H$ at a low level it is necessary to decrease the time step when increasing $\eta_0$.

Finally we apply a \textit{clipping} condition in order to prevent any spurious energy injection coming from singular solutions of equations (\ref{E=0_alv}) and (\ref{H=eta0}). Indeed as the forcing $\bff^H$ is random in time we cannot prevent the value of $D$ given in (\ref{D}) to be zero and lead to singular solutions $a(\bk,t)$ and $b(\bk,t)$. In addition we do not want to force energy nor helicity in the dissipative range (corresponding to scales $k\ge k_{\nu}$ where $k_{\nu}\approx \varepsilon^{1/4}\nu^{-3/4}$). Therefore at each time step the helical forcing $\bff^H$ is applied provided the following condition is satisfied
\begin{equation}
D(\bk,t)\ge |\bk||\bu_k|^2\left(A+B\left(\frac{k}{k_{\nu}}\right)^{\beta}\right),
\end{equation}
where $A$, $B$ and $\beta$ are positive constants whose values depend on $\eta_0$.

\bibliography{ref}
\end{document}